\documentclass{ieeeaccess}
\usepackage{cite}
\usepackage{amsmath,amssymb,amsfonts}
\usepackage{algorithmic}
\usepackage{graphicx}
\usepackage{textcomp}
\usepackage{sidecap}
\usepackage{multicol}
\usepackage{balance}
\usepackage{hyperref}
\usepackage{soul}
\usepackage{longtable}
\usepackage{array}
\usepackage{ragged2e}
\usepackage{makecell}
\usepackage{amsthm}
\usepackage{amsfonts}
\usepackage{color}
\usepackage{soul}
\usepackage{tabularx}
\usepackage{adjustbox}
\usepackage{adjustbox,lipsum}
\usepackage{rotating}
\usepackage{xcolor,colortbl}
\usepackage{subfiles}
\usepackage{multicol}
\usepackage{multirow}
\usepackage{stfloats}
\usepackage{url}
\usepackage{xurl}
\definecolor{Gray}{gray}{0.9}
\definecolor{accessblue}{cmyk}{1, 0.3, 0, 0.2}
\definecolor{greycolor}{cmyk}{0,0,0,.8}
\def\BibTeX{{\rm B\kern-.05em{\sc i\kern-.025em b}\kern-.08em
    T\kern-.1667em\lower.7ex\hbox{E}\kern-.125emX}}

\usepackage[font={sf,small,stretch=0.84},
            labelfont={bf,color=accessblue}
            ]{caption}  

\begin{document}
\history{Date of publication xxxx 00, 0000, date of current version xxxx 00, 0000.}
\doi{xx.xxxx/ACCESS.xxxx.DOI}

\title{From Assistive Technologies to Metaverse – Technologies in Inclusive Higher Education for Students with Specific Learning Difficulties: A Review}
\author{\uppercase{Gokul Yenduri}\authorrefmark{1} \IEEEmembership{Student~Member, IEEE}, \uppercase{Rajesh Kaluri}\authorrefmark{1}, \uppercase{Dharmendra Singh Rajput}\authorrefmark{1}, \uppercase{Kuruva Lakshmanna}\authorrefmark{1}, \uppercase{Thippa~Reddy~Gadekallu}\authorrefmark{1,2,4,5} \IEEEmembership{Senior~Member, IEEE},  \uppercase{Mufti Mahmud}\authorrefmark{3}, \uppercase{David J. Brown}\authorrefmark{3}
}

\address[1]{School of Information Technology and Engineering, Vellore Institute of Technology, Tamil Nadu, India (e-mail: gokul.yenduri, rajesh.kaluri, dharmendrasingh, lakshman.kuruva {@vit.ac.in})}
\address[2]{Department of Electrical and Computer Engineering, Lebanese American University, Byblos, Lebanon (e-mail: thippareddy{@ieee.org})}
\address[3]{Nottingham Trent University, Clifton Lane, Nottingham, UK (e-mail: mufti.mahmud, david.brown{@ntu.ac.uk})}
\address[4]{Zhongda Group, Haiyan County, Jiaxing City, Zhejiang Province, China, 314312}
\address[5]{College of Information Science and Engineering, Jiaxing University , Jiaxing 314001, China}

\corresp{rajesh.kaluri@vit.ac.in, thippareddy@ieee.org}

\begin{abstract}
The development of new technologies and their expanding use in a wide range of educational environments are driving the transformation of higher education. Assistive technologies are a subset of cutting-edge technology that can help students learn more effectively and make education accessible to everyone. Assistive technology can enhance, maintain, or improve the capacities of students with learning difficulties. Students with learning difficulties will be greatly benefited from the use of assistive technologies. If these technologies are used effectively, students with learning difficulties can compete with their peers and complete their academic tasks. We aim to conduct this review to better understand the role of assistive technologies in providing inclusive higher education for students with learning difficulties. The review begins with the introduction of learning difficulties and their causes; inclusive education and the need for assistive technologies; the reasoning for conducting this review; and a summary of related reviews on assistive technologies for students with learning difficulties in inclusive higher education. Then, we discuss the preliminaries for the learning difficulties type and assistive technology. Later, we discuss the effects of assistive technology on inclusive higher education for students with learning difficulties. Additionally, we discuss related projects and support tools available in inclusive higher education for students with learning difficulties. We also explore the challenges and possible solutions related to using assistive technology in higher education to provide inclusive education for students with learning difficulties. We conclude the review with a discussion of potential promising future directions.
\end{abstract}

\begin{keywords}
Learning Difficulties, Inclusive Higher Education, Assistive Technologies. 
\end{keywords}

\titlepgskip=-15pt

\maketitle

    \begin{table*}[h!]
    \renewcommand{\arraystretch}{1}
    \caption{List of key acronyms.}
 \centering
    \begin{tabular}{|l|l|}
        \hline
        \textbf{Acronyms} & \textbf{Description}\\
        \hline
        {ADD}    & {Attention Deficit Hyperactivity Disorder}\\
        \hline
        {AI}    & {Artificial Intelligence}\\
        \hline
        {AR}    & {Augmented Reality}\\
        \hline
        {ARC}    & {Augmented Classroom}\\
        \hline
        {ATIA}    & {Assistive Technology Industry Association}\\
        \hline
        {BCI }    & {Brain Computer Interface}\\
        \hline
        {CAPD}    & {Central Auditory Processing Disorder}\\
        \hline
        {DT}    & {Digital Twin}\\
        \hline
        {DCD}    & {Developmental Co-ordination Disorder}\\
        \hline
        {HMD }    & {Head Mount Display}\\
        \hline
         {HCI}    & {Human-Computer Interaction }\\
        \hline
        {IoT}    & {Internet of Things }\\
        \hline
        {IEEE}    & {Institute of Electrical and Electronics Engineers }\\
        \hline
        {MHD}    & {Mental Health Disorders }\\
        \hline
        {NDD}    & {Neuro Developmental Disorder}\\
        \hline
         {NVLD}    & {Nonverbal Learning Disorder}\\
        \hline
        {NGSS}    & {Next Generation Science Standards}\\
        \hline
        {PD}    & {Psychological Disorders}\\
        \hline
        {PSO }    & {Particle Swam Optimization}\\
        \hline
        {SVM }    & {Support Vector Machine}\\
        \hline
        {UNESCO}    & {United Nations Educational, Scientific and Cultural Organization}\\
        \hline
        {VR}    & {Virtual Reality}\\
        \hline
        {WHO}    & {World Health Organization}\\
        \hline
        {WIPO }    & {World Intellectual Property Organization}\\
        \hline
        {XR}    & {Extended Reality}\\
        \hline
         {3D}    & {3Dimension}\\
         \hline
    \end{tabular}
    \label{Tab:acronym}
\end{table*}
\section{Introduction} 

A learning difficulty is any abnormality of the body or mind that hinders a person's ability to do certain tasks and interact with the outside world. According to the World Health Organization (WHO), 15\% of the global population is disabled, of whom 2\% to 4\% have significant learning difficulties. According to the WHO, this global estimate of learning difficulties is rising due to an ageing population, the rapid spread of chronic diseases, and advances in the methodologies used to diagnose learning difficulties. \cite{WHO}. According to a UNICEF study, around 240 million children worldwide are facing issues with learning difficulties. Most assessments of child well-being indicate that children with learning difficulties have a worse quality of life than children without learning difficulties \cite{UNICEF}. Students with learning difficulties may suffer from emotional, mental, physical, or developmental problems. Education will provide students with learning difficulties with a feeling of self-worth. Learning difficulties are an underrated but major element in educational discrimination. Students with learning difficulties are among the most underserved categories regarding adequate higher education. All students with learning difficulties must receive a proper inclusive education for their growth and social development.

\subsection{Causes of Learning Difficulties}
With the assistance of modern technologies, researchers could pinpoint all likely causes of learning difficulties. Some learning difficulties are caused by prenatal and neonatal hazards, psychological and physical stress, and environmental exposure. Several risk factors are present at birth and run in families, according to recent studies \cite{Understanding}. Therefore, there is an increased risk of learning difficulties in children of those with learning difficulties. To better understand learning difficulties, it is essential to investigate how students' brains adapt to reading, writing, and solving mathematical problems \cite{schnepel2022systematic}. More individuals with learning difficulties suffer from physical and psychological illnesses. There are curable conditions associated with a person's learning difficulties. Autism, Attention Deficit Hyperactivity Disorder, Schizophrenia, Mania, and Pica are a few of the conditions that may lead to learning difficulties \cite{mizen2012learning}. Problematic behaviour may also indicate underlying mental or physical health issues. Additionally, family members, educators, and caretakers must assist the student in overcoming these obstacles \cite{khalil2022eating}. Students with learning difficulties have fewer opportunities to get an inclusive technical education, but this barrier may be overcome with the use of assistive technology. Numerous studies have identified methods for assisting individuals with learning difficulties.

\subsection{Inclusive Education for Students with Learning Difficulties}
Traditionally, students with learning difficulties were seen as inferior and barred from normal classes due to their cognitive disorders. Then, they were educated at specialized institutions of higher learning. Inclusive education allows for integrating students with learning difficulties into regular classrooms with their normally developing classmates and helps them to face real-world problems \cite{badilla2020augmented}. 1994's Declarations of Salamanca on Special Education outlined the ideas of inclusive education. The declaration emphasizes with a commitment to education for all students, acknowledging the need and urgency of providing education to all students, adolescents, and adults within the regular education system \cite{salamanca}. Regular education with an inclusive focus is the most effective strategy for eradicating discriminatory attitudes, developing welcoming communities, building an inclusive society, and achieving education for all \cite{aly2021abet}. In addition, inclusive education will provide an effective education for most students, hence boosting the efficiency and cost-effectiveness of the education system as a whole \cite{meijer2019financing}.

\subsection{Need of Assistive Technologies for Students with Learning Difficulties in Higher Education}
 Students with different types of learning difficulties confront various obstacles in higher education. Students with learning difficulties are excluded from school, society, and mainstream development programs due to a lack of crucial support and fair participation opportunities \cite{hayes2017disabilities}. If technology is implemented properly, students with learning difficulties will be able to participate in the general education curriculum because they will have access to simpler and more adaptable methods of completing their tasks. It is vital to pick assistive technology based on a student's requirements, not their difficulties category. As assistive technology is intended to accomplish a certain objective, it is crucial to choose the appropriate tool for the task \cite{martiniello2022exploring}. Assistive technology may include hardware, software, and peripherals that aid students with learning difficulties in completing their assignments \cite{soetan2021attitude}. Utilizing assistive technology in higher education will aid students with learning difficulties in remaining competitive with their peers, fostering social engagement, boosting self-confidence, and enhancing academic success.
 
\subsection{Motivation}
\begin{figure*}[ht!]
\centering
\includegraphics[width=.5\linewidth]{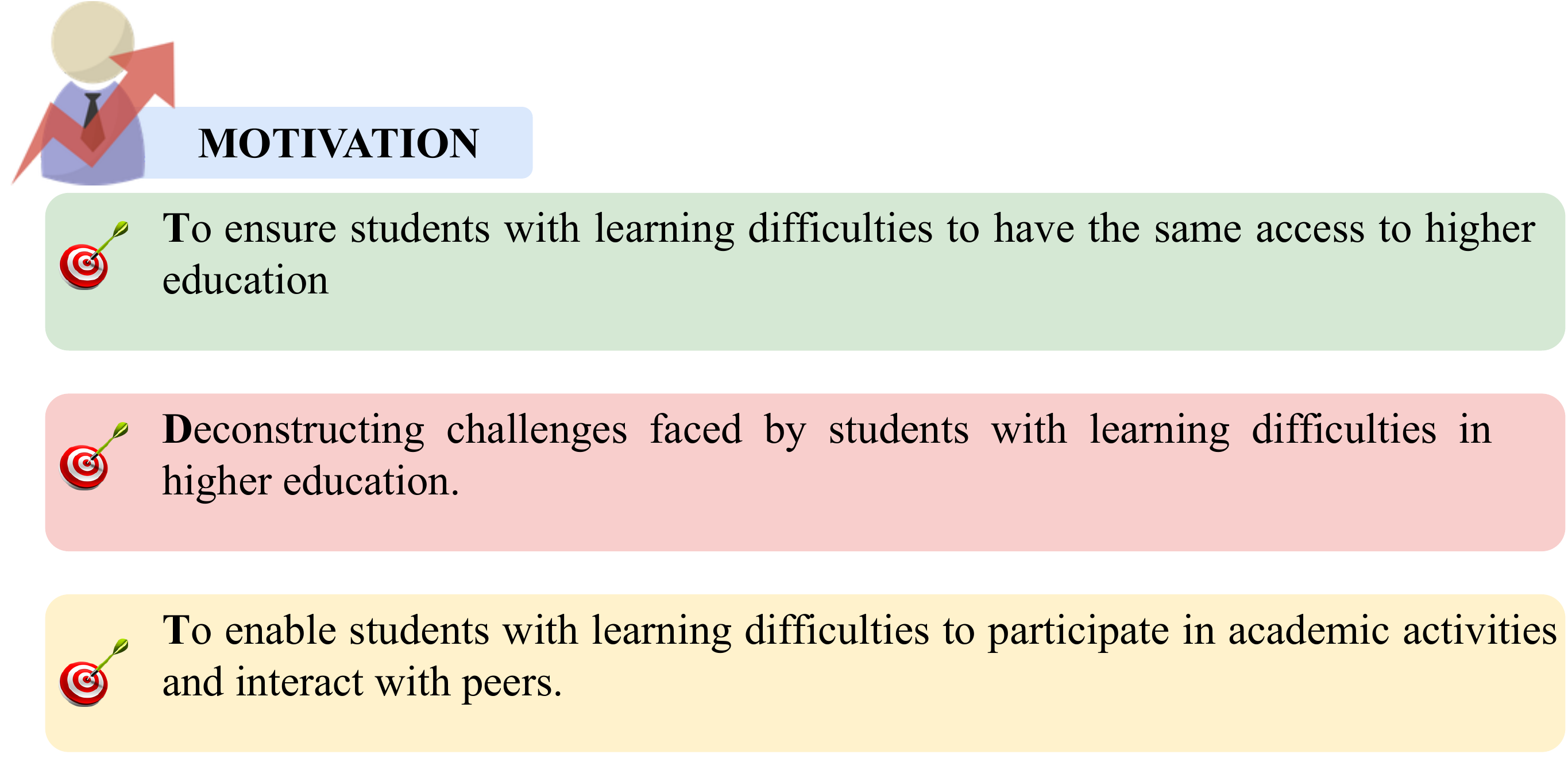}
\caption{Motivation for Review}
\label{MOTO}
\end{figure*}

In this section, we discuss the motivation for conducting this review, and the motivation for doing this review is depicted in Fig.\ref{MOTO}. Our study is limited to those assistive technologies that make it possible for students with learning difficulties to overcome the challenges they face while competing  with their peers. Our work is restricted in the case of students with physical impairments, as these assistive technologies reviewed in our study may not address all physical disabilities, which require a separate focus. These assistive technologies for physical impairments not only need to focus on alternative or customized devices but also need to address the student's learning difficulties along with their physical impairments as a whole.

\subsubsection{To ensure Students with Learning difficulties to have the same access to Higher Education}
Students with learning difficulties, such as dyslexia or autism, struggle to adapt to the conventional learning environment. They have difficulty following instructions, course content, and even their textbooks. They need individualized instruction and cannot withstand the pressure of competition. Moreover, these students are often bullied by their classmates, which may further demoralize them \cite{macconville2007looking}. Innovative assistive technology and a personalized learning environment can help students get past these problems and get ready to compete with their classmates and do well in academics \cite{kouatli2020need}.

\subsubsection{Deconstructing Challenges faced by the Students with Learning Difficulties in Higher Education}
Academically struggling students with learning disabilities may experience stress \cite{zeeni2018media}. Possible stressful sentiments of low self-worth may arise as a result. Many students may not perform in the classroom because they fear making some mistakes in their work \cite{kohli2018specific}. Those with learning difficulties with self-determination and the help of assistive technology can overcome the stigma of having a learning difficulty and deal with the problems of higher education effectively \cite{field2003self}.

\subsubsection{To enable Students with Learning Difficulties to participate in Academic Activities and interact with Peers}
The use of assistive technology will provide students with chances for experiential learning. Students can engage in independent academic activities or work collaboratively with other students instead of idly waiting for assistance \cite{pilgrim2012new}. The use of assistive technology allows students with learning difficulties to learn at their own pace. This self-paced learning results in less pressure and improved communication skills, attention, and behaviour \cite{parette2008benefits}. Assistive technology can help students with learning difficulties engage more readily in cooperative learning activities, as students with learning difficulties may not possess either the needed academic or collaborative skills to participate fully \cite{bryant1998treatment}. With the help of assistive technologies students’ gaining independence as they were able to complete written assignments with minimal or no assistance \cite{bouck2010pen}.   

\begin{table*}[t]
    \renewcommand{\arraystretch}{1.15}
	\caption{Summary of Related Reviews on Assistive Technologies for Students with Learning Difficulties in Inclusive Higher Education.}
	\label{Table:Summary_ExistingSurveys}
	\centering
	\resizebox{\textwidth}{!}{%
	\begin{tabular}{|p{1.30cm}|p{7.25cm}|p{6.25cm}|}
		\hline 
		\multirow{1}{*}{\textbf{Ref}} & \multirow{1}{*}{\textbf{Contributions}} & \multirow{1}{*}{\textbf{Limitations}} \\
		\hline
		\hline
		
		\multirow{1}{*}{\cite{svensson2021effects}} & This study investigated the effectiveness of assistive technology for students with learning difficulties. & This study limits the most recent technical developments in assistive technology, which can help in inclusive higher education. \\ \hline
		\multirow{1}{*}{\cite{reddy2021assistive}} & This study explored the impact of assistive technology on mathematics in higher education. & This study did not provide insights into how this assistive technology can help students with learning difficulties in inclusive higher education.  \\ \hline
		\multirow{1}{*}{\cite{barua2022artificial}} & This study explored the role of artificial intelligence in personalized assistive technology for neurodevelopmental disorders students in their education.  & This study did not address the problems associated with incorporating AI as an assistive technology for students with learning difficulties in inclusive higher education.   \\ \hline
		\multirow{1}{*}{\cite{farhan2022comparative}} & This study compared an assisted e-learning interface amongst students with and without visual and auditory impairments. & This study did not address the difficulties in adapting the e-learning interface for students with visual and hearing difficulties.  \\ \hline
		\multirow{1}{*}{Our work} & Our work explored the role of assistive technologies in providing inclusive education for students with learning difficulties in higher education & -- \\ \hline
		
	\end{tabular}
	}
\end{table*}

\subsection{Related Works and Contributions}
Several researchers have worked on assistive technology for students with learning difficulties in inclusive higher education. The overview of these studies is provided in Table \ref{Table:Summary_ExistingSurveys}.

To assist reading and writing, tablets featuring text-to-speech and speech-to-text capabilities have been launched in recent years \cite{dawson2019assistive}. Since the 1980s, the challenges they face in written and spoken language, arithmetic, reasoning and memory that are the result of their learning difficulties have been mitigated with the use of various assistive technology. However, few scientific studies have examined the advantages of this method. Idor Svensson et al. evaluated the influence of assistive technology on students with learning difficulties. In their research, 149 individual students participated. The intervention group was provided with 24 training sessions in assistive technology, whereas the control group received routine care. In a single year, the intervention and control groups attained the same degree of improvement as the normative population. However, neither immediately after the intervention nor one year later did gains vary across groups. They found that the use of assistive technology improved reading skills, especially for students with learning difficulties. It was also observed that the intervention boosted motivation leading to task completion. Furthermore, their research demonstrated the challenges of assessing individuals with learning difficulties in terms of their capacity to comprehend and interpret information  \cite{svensson2021effects}.

The landscape of higher education is constantly changing as a result of the quick adoption and spread of new technologies in teaching and learning methods \cite{kompen2019personal}. Assistive technologies, which include a variety of specialised tools, are used to help students access education and participate freely and actively in the learning process, which improves learning and promotes the educational system \cite{bugaj2022new}. Pritika Reddy et al. examined students' perceptions of the use of various assistive technologies, including mobile learning, tablet learning, lecture capture, gamification, and online intelligent systems for learning and student assistance in higher education in mathematics teaching at the university level. They also assessed the opinions of assistive technology among mathematics students in online mode. Their findings concluded that assistive technologies helped students with learning disabilities to understand mathematics better, and the students showed a positive outlook on the use of these technologies in mathematical education \cite{reddy2021assistive}.

Neurodevelopmental disorders (NDDs), which also include developmental disabilities and specific learning difficulties like attention deficit hyperactivity disorder (ADHD), dyslexia, and autism spectrum disorders (ASD), as well as a wide range of mental health disorders (MHDs), including stress, anxiety, psychotic disorders, and severe depression, are frequently associated with psychological disorders (PDs) that first emerge in adolescence or early childhood. Over the last 20 years, there have been notable increases in the diagnosis of mental diseases on a global scale as well as rapid growth in the rate of a number of mental health problems. Depending on the kind of PD, students may struggle with socialisation, communication, and adapting to changes in their environment, which could make it difficult for them to concentrate effectively. To improve outcomes for students, treatment has to be carried out quickly and effectively \cite{ogundele2022classification}. In order to address learning difficulties in students with a variety of NDDs, Prabal Datta Barua et al. examined the complexity and effectiveness of AI-assisted solutions created using machine learning models. Their study provided a summary of the data showing how AI technology may be utilised to improve social interaction and support training. They concluded that AI solutions aren't completely effective at resolving the problems associated with learning difficulties. They recommended that in the future, AI technologies may be improved with an emphasis on assisting students with NDDs \cite{barua2022artificial}.


In summary, researchers used a variety of technology to assist students with a variety of learning difficulties. To our knowledge, no research has been undertaken on the usage of different assistive technologies to aid higher education students with learning difficulties which provided the motivation for this review. This research examines the role that assistive technology in providing inclusive education to students with learning difficulties in higher education.

\subsection{Systematic Literature Review}
 The following phases constitute the literature review that was used in this study to investigate the role of assistive technology for students with learning difficulties in higher education. First, we discuss the shortcomings of previous review articles and the reasons for conducting this study. Investigating relevant scientific and research publications on the use of assistive technology for impaired students in higher education is the next step. We place a strong emphasis on peer-reviewed, high-quality papers published in reputable books,  conferences, seminars, symposiums, and journals. The references utilized in this study were found on well-known archive services including Google Scholar and arXiv as well as well-regarded publications like Springer Nature, Wiley, Elsevier, Taylor and Francis, MDPI, and IEEE. Additionally, the keywords AI, XR, computer vision, the metaverse, HCI, and digital twins are utilised to identify relevant references and publications about assistive technology for learning difficulties in inclusive higher education such as Dyslexia, Dyspraxia, Dyscalculia, Dysgraphia, Auditory Processing Disorder, Visual Processing Disorder, Nonverbal Learning Disorder, and Apraxia of Speech. The retrieved articles are all screened based on their titles in the next phase. We didn't include any papers with poor-quality material from predatory journals. After that, we reviewed the abstracts of the papers to determine their contributions. The data needed for our analysis of the use of assistive technology for students with learning difficulties in higher education is extracted in the final step \cite{kitchenham2009systematic}.

\subsection{Paper Organization}

\begin{figure*}[ht!]
	\centering
	\includegraphics[width=0.8\linewidth]{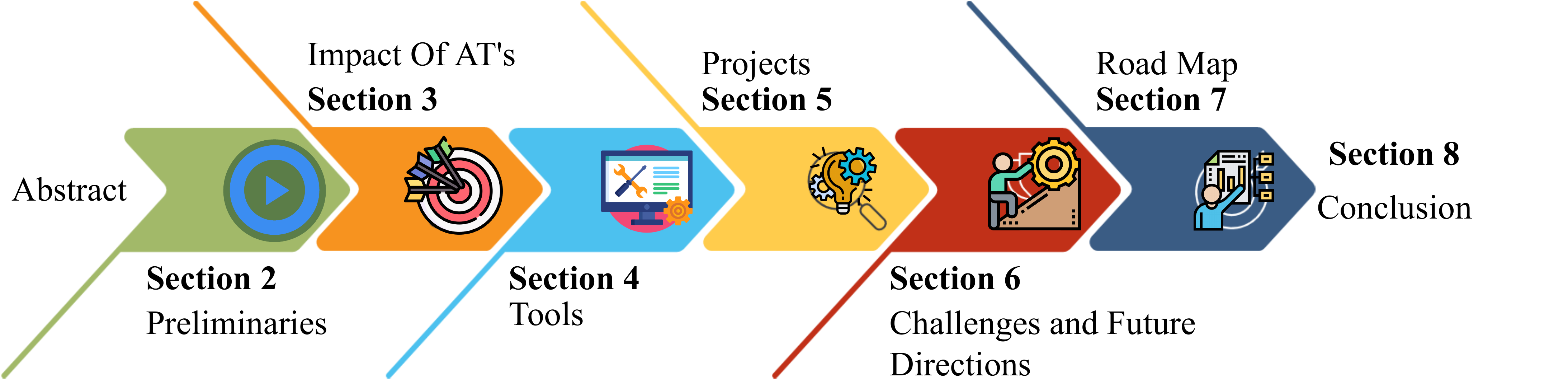}
	\caption{The Schematic Organisation of the Role of Assistive Technologies in Providing Inclusive Education for Students with Learning Difficulties in Higher Education}
	\label{Outline of the paper}
\end{figure*}

    Section II presents the preliminaries of assistive technologies and types of learning difficulties. In Section III, we discuss the impact of assistive technologies in providing inclusive education for students with learning difficulties in higher education, which includes AI, XR, the metaverse, HCI, and digital twins. Then we discuss the projects working toward inclusive education for students with learning difficulties in higher education in Section IV. Next, Section V is an overview of assistive technology tools. To drive further studies on assistive technologies for learning difficulties students in higher education, in section VI, we discuss challenges in integrating assistive technologies for slow learners in higher education and future directions. Section VII is the road map. Finally, we conclude the paper in Section VIII. For clarity, the organisation of this review is presented in Fig. \ref{Outline of the paper}, and a list of frequently used acronyms is listed in Table~\ref{Tab:acronym}.

\section{Preliminaries}
\label{Sec:Preliminaries}
This section provides an overview of learning difficulties, which are identified based on a systematic literature review, followed by a discussion of assistive technology for learning difficulties in higher education.
\subsection{Types of Learning Difficulties}
\begin{figure*}[ht!]
	\centering
	\includegraphics[width=1.0\linewidth]{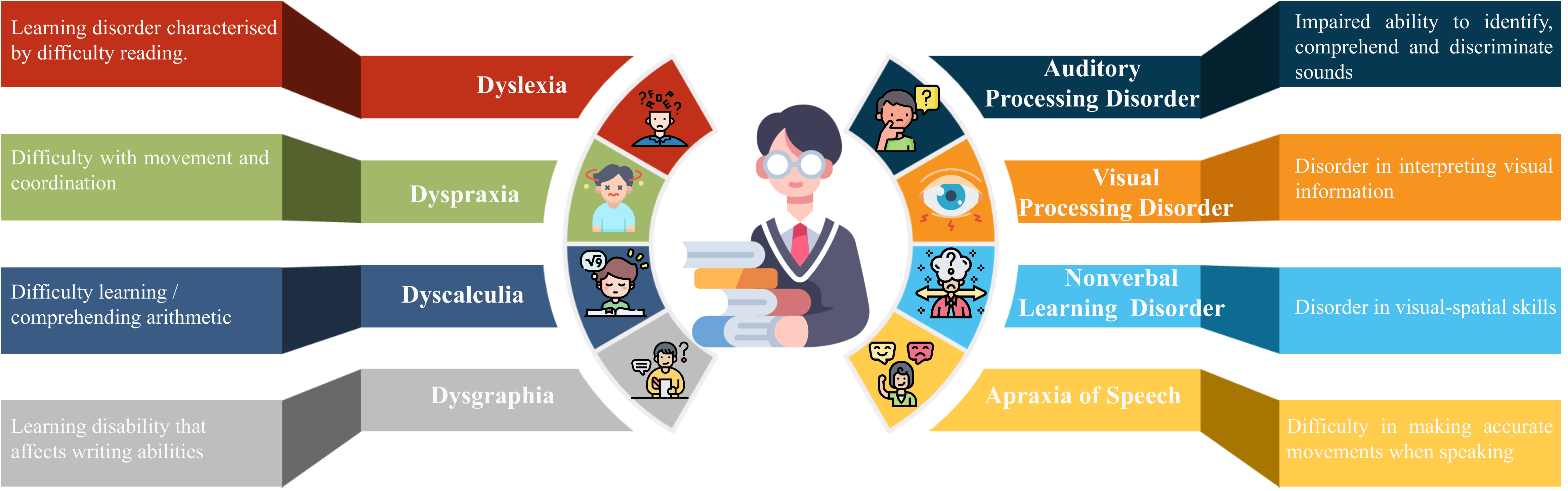}
	\caption{Types of Learning Difficulties}
	\label{Learning Disabilities}
\end{figure*}

The type of various learning difficulties and their effects are depicted in Fig. \ref{Learning Disabilities}

\subsubsection{Dyslexia}
Dyslexia is a learning disorder. Identifying speech sounds, reading, and decoding letters and words can be challenging for students with dyslexia. Students with dyslexia may also have trouble speaking and expressing themselves, and their ideas \cite{usman2021advance}. They may also find it hard to organise their thoughts during talks. Despite the impairment in language processing regions of the brain caused by dyslexia, students with dyslexia can compete with their peers with the aid of assistive technologies and appropriate intervention.
\subsubsection{Dyspraxia}
Dyspraxia is also known as developmental coordination disorder or DCD. Dyspraxia is a motor disorder based on the brain. It influences large and fine motor skills, motor planning, and coordination \cite{o2019children}. Despite the fact that it can influence cognitive skills, it is unrelated to intelligence. A student with dyspraxia has difficulty with movement and coordination. Students may have difficulty handling objects and may also tend to bump into things. The student may also have trouble speaking, be sensitive to light, touch, taste, or smell, and have trouble moving his or her eyes.
\subsubsection{Dyscalculia}
Dyscalculia makes it challenging to understand, and process arithmetic \cite{cappelletti2014commonalities}. In addition to counting and simple mental math difficulties, students have trouble telling time and following directions. The following abilities are affected by dyscalculia: problems with comprehending how numbers operate and their relationships, mathematical problem-solving difficulties, difficulty with learning basic calculations, and they also face difficulty in compiling and documenting data.

\subsubsection{Dysgraphia}
The neurological condition known as dysgraphia impairs writing and fine motor skills. It is a learning difficulty that affects almost every element of writing, including spelling, legibility, word size, and expression of a tense grip that might result in a hurting hand.  Bad spatial planning, inconsistent writing, poor spelling and missing or incomplete words are all symptoms of dysgraphia \cite{dimauro2020testgraphia}.

\subsubsection{Auditory Processing Disorder}
It is also known as Central Auditory Processing Disorder (CAPD) because it affects a person's ability to detect, understand, and identify sounds while having normal hearing. Significant difficulty understanding speech, particularly in noisy environments; difficulty following multi-step spoken instructions delivered without visual aids; distraction by loud or unexpected sounds; difficulty paying attention to lengthy lectures or other extended listening sessions; difficulty remembering and/or efficiently summarising verbally delivered information; and difficulty reading, spelling, and/or writing are some of the symptoms of CAPD \cite{moore2018auditory}.
\subsubsection{Visual Processing Disorder}
A student with Visual Processing Disorder(VPD) has difficulty comprehending visual information. The student may struggle with reading or distinguishing between similar-looking objects. Those with visual processing problems may experience difficulties with hand-eye coordination. Students with visual processing disorder face difficulties like Confuse words that look similar, reverse letters or numbers, Lacking adequate reading comprehension, make copying errors and frequently forgetting letters, numbers, and words, being bad spellers, having uneven or poorly spaced handwriting and having difficulty in understanding multi-step instructions, and have trouble telling time and comprehending the idea of time \cite{janarthanan2017visual}.

\subsubsection{Nonverbal Learning Disorder}
The most underdiagnosed, misunderstood, and ignored learning impairment is nonverbal learning disorder (NVLD). Impaired visual, spatial, and organisational skills, difficulty recognising and interpreting nonverbal signals, and poor motor function are all symptoms of the neurological condition. The symptoms of NVLD also include problems related to social interactions, reading nonverbal signs, understanding facial expressions, using appropriate language in social situations, coordination of the body, and fine motor abilities. Students affected by NVLD may face difficulties with organisation, planning, and concentration, as well as reading comprehension and writing expression at a higher educational level \cite{banker2020spatial}.

\subsubsection{Apraxia of Speech}
A student with apraxia of speech struggles to talk clearly and make appropriate gestures. In apraxia of speech, the speech muscles are not weak. Rather, the brain has problems directing and/or coordinating the motions, so the muscles do not function appropriately. Apraxia in students may cause difficulty in imitating and producing sounds on their own, may add new sounds, omit sounds, or pronounce sounds incorrectly, and may pronounce something correctly one time and incorrectly the next \cite{botha2019primary}.

\subsection{Assistive Technologies}
Anything software, hardware or peripherals that aid students with learning difficulties in overcoming their educational obstacles and developing new skills fall under the umbrella of assistive technology. Students with learning difficulties need assistive technology to improve their abilities. They will be able to receive a high-quality education on par with their peers using assistive technology.\\
\textbf{Definition 1:} 
According to WHO, the systems and services involved in providing assistive products and services are collectively referred to as assistive technology. \cite{Assistivetechnology}.\\
\textbf{Definition 2:} According to the Assistive Technology Industry Association (ATIA), any tool, piece of equipment, piece of software, or product used to enhance, maintain, or strengthen the functional capacities of individuals with disabilities is known as assistive technology. \cite{Assistivetechnology2}.\\
\textbf{Definition 3:} The federal defines assistive technology as any tool, apparatus, or system, whether purchased commercially off-the-shelf, adapted, or customized, that can be used to enhance, maintain, or improve the functional capacity of people with impairments \cite{Assistivetechnology3}.\\ 
\textbf{Definition 4:}  According to the United Nations Educational, Scientific and Cultural Organization (UNESCO), anything that is utilized to enhance, maintain, or improve the functional capacities of people with impairments is considered assistive technology \cite{Assistivetechnology4}. \\
\textbf{Definition 5:} According to the Institute of Electrical and Electronics Engineers (IEEE), anything that aids a person in achieving increased performance, function, or quicker access to information is considered assistive technology. \cite{Assistivetechnology5}. \\
\textbf{Definition 6:} According to The European Accessibility Act (EU, 2019), assistive technology refers to any device, appliance, service, or combination of processes, including computer programs, that is used to maximize, maintain, replace, or enhance the functional skills of people with disabilities. \cite{Assistivetechnology8}. \\  
\textbf{Definition 7:} According to the International Organisation for Standardisation's (ISO9999(2022)) standard on assistive products, assistive technology is any item that was produced with a specific focus on serving the needs of people with disabilities or that is generally available and used by or for people with disabilities \cite{Assistivetechnology7}. \\
\textbf{Definition 8:} According to the World Intellectual Property Organization (WIPO), the term "assistive technology" refers to a broad range of technologies and goods, from relatively simple gadgets like a walking stick or reading glasses to sophisticated, high-tech systems like assistive robots or software that recognizes gestures or emotions. \cite{Assistivetechnology8}. \\

\section{ The Significance of Assistive Technology in Higher Education in Ensuring Inclusive Education for Individuals with Learning Difficulties}
\begin{figure*}[ht!]
	\centering
	\includegraphics[width=1.0\linewidth]{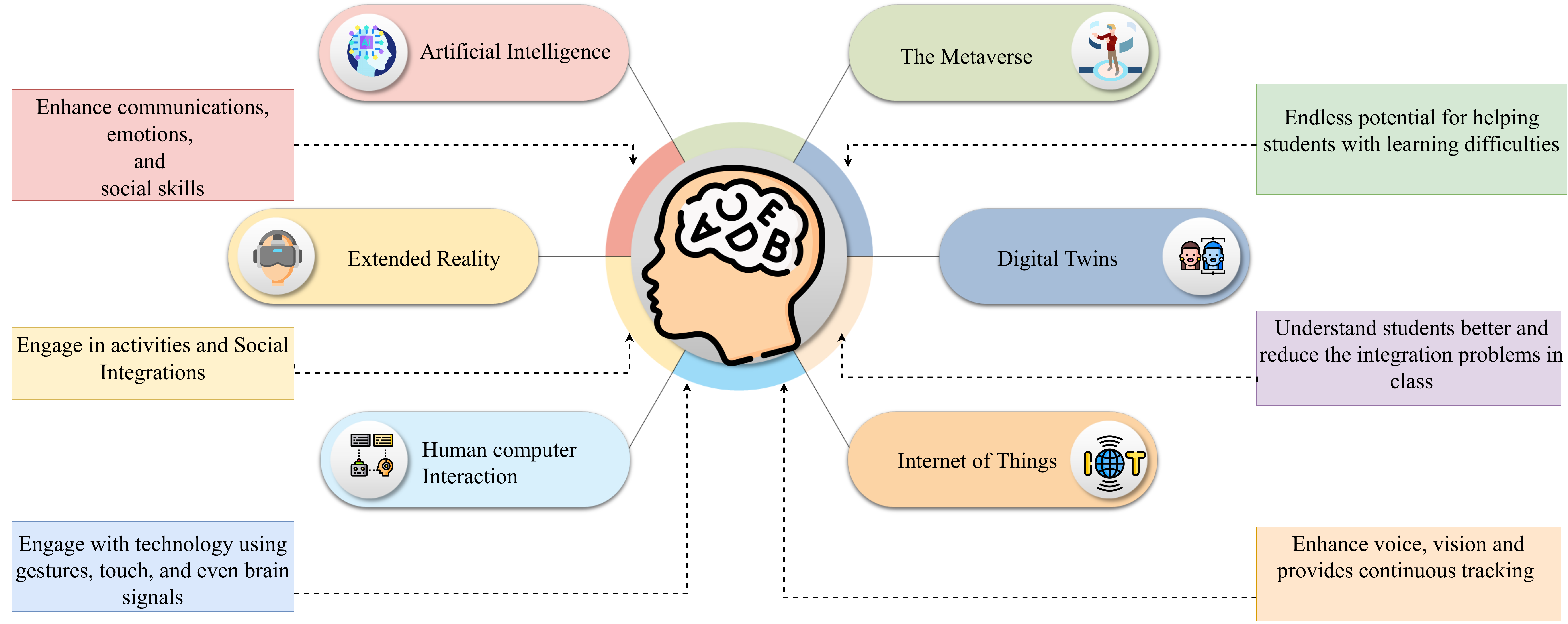}
	\caption{Assistive Technologies for Students with Learning Difficulties in Higher Education}
	\label{AT}
\end{figure*}

Based on the systematic literature review, it is understood that the development of assistive tools in future will be heavily reliant on contemporary technologies like AI, XR, IoT, HCI, digital twins, and the metaverse. There are numerous studies that concentrate on lower-end assistive technology for learning difficulties \cite{tarhini2019information}. To the best of our knowledge, there is no review that addresses all of these cutting-edge technologies for assisting students with learning difficulties.
In this section, we address the role of AI, XR, IoT, HCI, digital twins, and the metaverse as assistive technologies in inclusive education that assist students to overcome their learning difficulties and Fig. \ref{AT} depicts the assistive technologies covered in this study. 
\label{Sec:Technical}
\subsection{Artificial Intelligence} 

Artificial intelligence (AI) represents significant advances in computer science and data processing that are rapidly revolutionizing numerous sectors \cite{tarhini2022artificial}. AI refers to the simulation of human intelligence in machines that are programmed to replicate human thought and behavior \cite{ullrich2022data}. The use of AI makes inclusive education a reality. Students with learning difficulties can be integrated into regular classrooms and educated alongside their peers \cite{br2022ontology}. AI is advancing rapidly in the education sector and reducing the gap between students and teachers. 
Learning-difficulty-specific cognitive systems are also being developed by researchers using AI \cite{ferenczy2022speech}. Google and Microsoft are developing AI-based tools such as the immersive reader and google docs that can assist students with learning difficulties. The early identification of students with learning difficulties can be facilitated by AI \cite{movaghar2022advancing}. AI can assist students in enhancing their reading comprehension by reading the text aloud and also provide comprehensive feedback on written work. AI enhances reading fluency in reading for students with learning difficulties. Students with autism struggle with both verbal and nonverbal communication. Social skill development can be difficult for them. To solve this problem, QTrobot was created. This humanoid robot was developed to teach social skills to autistic children. The robot NAO and the virtual assistant Siri are two more examples that help students with autism spectrum disorder(ASD) learn social skills. Tools like ActiveMath employs techniques of AI to allow students greater flexibility in finding a convenient learning environment. The AI-enabled Widex's Evoke will help students in improving their hearing capabilities. Grammarly is an AI-based writing assistant which will improve the writing skills of students with learning difficulties. 

AI can help in the early detection of learning difficulties. This can help students understand their condition and prepare for future challenges. A. Jothi Prabha et al. created an eye movement analysis model for the detection of dyslexia. They used an eye tracker to observe eye movement. The eye movement data includes fixations, saccades, transients, and distortions of the participants. Principal component analysis was used to identify high-level properties from raw eye tracker data. For the diagnosis of dyslexia in students, a PSO-based Hybrid Kernel (SVM-PSO) was created. Their results showed the proposed model's prediction accuracy was 95\% more accurate than that of the Linear SVM model. The model was validated over 187 individuals. They concluded that with the use of eye movement data and machine learning, the development of very precise prediction models was possible. Their method was also offered as a screening tool for dyslexia diagnosis \cite{jothi2022prediction}.

Students with learning difficulties have more emotional and behavioral challenges in the classroom than their peers without learning problems. In order to understand this issue, Nihal Ouherrou et al. conducted research on the benefits of using information and communication technology to understand the emotional factors of students with learning difficulties in online classrooms. In order to analyse the effects of the virtual learning environment, 42 students were divided into two groups. They considered seven basic facial expressions (fear, anger, disgust, surprise, happiness, neutral, and sorrow ) and assessed students' emotions using AI while they played a learning game. The results indicated that students with learning difficulties experience the same range of emotions as children without learning difficulties. Furthermore, they concluded that students with learning difficulties report fewer negative emotions compared to their peers in virtual learning environments \cite{ouherrou2019comparative}.

Tools and intelligent learning environments based on AI can be used to create successful individualised educational techniques for children with learning difficulties. 

Students with special needs are not as likely to use e-learning as students without special needs. Not much is known about the barriers and facilitators that cause this difference. The opinions of 21 teachers who took part in preliminary trials of an adaptive learning system based on multimodal affect recognition for students with learning disabilities and autism were gathered by Penny J. Standen et al. through focus groups and interviews. The adaptive selection of learning materials is driven by the system's multimodal detection of emotional state and scoring of performance. The teachers' thoughts on the possible effects of the system were summed up in five themes. These themes focused on learning, engagement, and factors that might affect adoption. These were how the system could change the way they taught, how it could affect how well students learned, how it could affect the relationships between teachers and students and between students, how easy it was to use, and how it could be set up. Even though the teachers who volunteered as testers were very interested, they pointed out barriers to adoption that needed to be fixed. Their finding showed how important it is for teachers and students to be involved in the process of design and development \cite{standen2021teachers}.

Phaedra S. analysed the historical development of intelligent learning environments. In addition, they reported the significant challenges that arise with the inclusion of intelligent learning environments. Furthermore, they examine a variety of novel strategies for addressing these issues, such as teacher training, the employment of instructional robots, and responsive systems. They concluded that AI in education is fast altering traditional views on teaching and learning. They also stated that traditional school and classroom models will change a lot over the next few years and decades as technology improves and spreads throughout educational institutions \cite{mohammed2019towards}.

Using multimodal sensor data and machine learning, Penelope J. Standen et al. found that learning is linked to three emotional states: engagement, frustration, and boredom. Then, they figured out how to present the learning content so that the learner stays in the best emotional state and learns as quickly as possible. 67 people between the ages of 6 and 18, who were their own controls, took part in a series of sessions using the adaptive learning system so that it could be evaluated. Sessions alternated between using the system to choose the learning content based on how the learner felt and how well they learned (intervention) and just on how well they learned (control). Lack of boredom was the state with the strongest link to achievement, with both frustration and engagement positively related to achievement. They concluded that the intervention sessions were much more interesting and less boring than the control sessions, but the amount of work done did not change much. Their results suggest that activities that match the needs and emotions of the learner do increase engagement and that the system promotes emotional states that help learning. They suggested that longer exposure is also needed to figure out what effect it has on learning \cite{standen2020evaluation}.

In summary, AI has the ability to transform education by creating cutting-edge teaching tools for students with special needs. In order to provide more intelligent solutions, the industry is expanding, demanding more study and cooperation between educators, app developers, and engineers. When AI is used as assistant in making decisions, it is challenging to explain the decisions made by AI because of its black-box nature. The students' personal data may be required for training these AI models, the protect the privacy of the students data is also a concern.

\subsection{Extended Reality} 
The combination of virtual environments, human-machine interactions, and wearable technologies is collectively referred to as extended reality (XR). The term "XR" encompasses both virtual reality (VR) and augmented reality (AR). Through the use of VR, users may interact with objects and other people in a manner that seems authentic. To add digital elements to a live, real-world situation, AR uses a smartphone, tablet, or headset. Mixed reality (MR) brings together the real and virtual worlds by using powerful computer technology, images, and input methods \cite{morimoto2022xr}.

VR and AR can assist students with learning difficulties in many ways in their academic endeavors \cite{mccord2022student}. The students with learning difficulties are immersed in a 3D world like CoSpaces Edu containing auditory, touch, smell, and gustatory inputs. Students can interact by wearing a head-mounted display (HMD) and a haptic glove or by using a regular desktop PC and VR software \cite{itani2021light}. By adding 3D effects to actual information, AR applications such as Google Lens allow users to remain unbiased observers and recognise the augmented effects. A VR environment such as Google Expeditions will allow students with learning difficulties to participate in activities that are unrestricted by their impairment and allows them to study in the most effective way possible. VR can also help promote positive views toward people with learning difficulties among their peers \cite{wagemaker2022susceptibility}. Individualised VR settings provide autistic children with the opportunity to learn social interaction and nonverbal cues \cite{boyd2018vrsocial}. Virtual environments or input stimulation are adaptable to student preferences \cite{standen2005virtual}. VR and AR can promote motivation, facilitate engagement, strengthen cognitive skills, and improve memory in students with learning difficulties . They can also improve communication skills, especially among students with hearing problems \cite{drigas2022virtual}. VR can help autistic students with social interaction \cite{parsons2011state}. AR can improve language through the use of games provided by applications like Assemblr \cite{haddick2022metahumans}. Narrator AR applications can inspire students with learning difficulties to improve their handwriting.

Students with learning difficulties have physical, mental, and communicative limitations \cite{karabulut2022exceptional}. Teaching individuals with learning difficulties requires a specific blend of methods and tools. VR is one of these technologies that will serve as an effective learning aid for students with learning difficulties. Arik Kurniawati et al. offered these students VR games. This game motivates students to acquire the items in a VR-based educational environment. The game helps students practise identifying, selecting, and pointing at things in response to visual and audio stimulation. Students with learning difficulties, autism, and learning difficulties participated in their studies. The results indicated that the game is completely simple and accessible for students with learning difficulties. They determined that all participants understood the instructions with little assistance \cite{kurniawati2019class}.

ADHD, sometimes known as hyperactivity among students, is a prevalent neurodevelopmental disorder. ADHD is characterised largely by hyperactivity, inattention, and behavioural impulsivity. Traditional treatments often rely on clinicans and parents who may observe and analyse a patient's behaviour using behavioural scales; however, these treatments are time-consuming and ineffective in measuring behaviour \cite{lai2022towards}. In their experiment, Yunchuan Tan et al. combined several sensor technologies, such as eye movement sensors and electroencephalography sensors (EEG), and used virtual reality technology to develop an assessment and diagnostic approach for ADHD. This system provided a virtual classroom environment and included a number of activities, including an audio exam. The continuous performance tasks and the Wisconsin card sorting test was used to evaluate the students' attention, ability to think abstractly, and cognitive capability. They introduced distracting elements to their experiment and analysed the test participants' attention. In order to assess the subject's sustained attention and attention shift, they combined their test results with physiological data such as head and eye movements and EEG \cite{tan2019virtual}. They concluded that their approach can improve the levels of concentration in students with learning difficulties.

Jorge Fernandez Herrero et al. suggest a concept and application for an immersive VR system with a head-mounted display to develop and teach the emotional and social abilities of students with autism spectrum disorders. They chose two groups of seven high-functioning ASD students with comparable intellectual aptitudes. On the first group, they used immersive virtual reality as a pedagogical tool to recreate virtual socialising contexts over the course of 10 sessions, using their own intervention design to address social and emotional competencies. As the control group, the second group is not subjected to any intervention throughout the intervention period. The degrees of adaption and the observed improvements suggested that immersive virtual reality in the format described is consistent with the sensory preferences and visuospatial abilities of the ASD children who participated in this research. They concluded that immersive VR can be utilised effectively as a teaching tool for children with ASD \cite{herrero2020immersive}.

In summary, XR can be used to successfully improve the skills and abilities of people with dyslexia, social anxiety, ADHD, linguistic impairments, physical or motor disabilities, Down syndrome, and cognitive deficits. The compatibility of these XR devices with other assistive technologies and devices remains a challenge due to their major design flaws and lack of standards. 

\subsection{Human-Computer Interaction} 

Human-computer interaction (HCI) is a field of study that looks at the design of computer technology and, in particular, the users interact with computers. In HCI, cognitive science principles and methods allow software engineering and the human aspects of computing systems to work well together. HCI includes using gesture, touch, and even affords interaction via brain signals \cite{sabie2022decade}.

Universal design for HCI promotes inclusive education that produces accessible products for all students, including those with learning difficulties. The products accommodate individual preferences and abilities and efficiently transmit essential information independent of environmental conditions or the user's sensory capabilities. They are manipulable, reachable, approachable, and usable regardless of body size, posture, or mobility \cite{moghaddam2019massive}. These design principles result in products that are compatible with assistive technology and are more usable for everyone \cite{jeanneret2022web}. Students with learning difficulties struggle with conventional forms of expression, such as writing, but can demonstrate their comprehension in a number of innovative HCI ways by using video or screen-casting tools including Clips, iMovie, Audacity, and others. HCI provides a readily accessible collaborative environment in which students may produce and share project-based and other related materials using online storage tools such as Dropbox. HCI also helps in the creation of student-response tools which improve student-teacher interaction. OneNote Web Clipper is a highly adjustable immersive reader that has options for changing the text size and line spacing, showing the parts of speech, and more. A student with a visual difficulty may use VoiceOver to describe what is on the screen using synthetic voice or braille (with a linked braille display). Students may draw thoughts with the aid of HCI visual tools like Draw.io and Popplet. These HCI tools with universal design will support inclusive education and assist students in overcoming their learning challenges. 

The intelligent math e-tutoring system developed by Zikai Alex Wen et al. attempts to eliminate students' negative emotional responses. The technology identifies potential negative emotional actions by evaluating gaze, touchscreen inputs, and reaction time. The program then employs intervention strategies to prevent students from becoming irritated. Formative research carried out with five instructors of students with learning difficulties helped to develop this design. Teachers believed that the establishment of these intervention strategies would benefit students with learning difficulties and stated that among the intervention strategies available to them, giving students 'brain breaks' is the newest and most beneficial. Additionally, the instructors proposed that the system could be customised to identify negative emotional responses, in order to assist students with learning difficulties \cite{wen2021intelligent}.

Brain-Computer Interface technology (BCI) is an important aspect of HCI. Bio-signals acquired by wearable sensors are attracting significant interest beyond the traditional medical arena, in new paradigms such as education \cite{putze2022understanding}. Attention is a bio-signal that may be detected and analysed using BCI technology by measuring the frequency of alpha (8-13 Hz) and beta (14-30 Hz) waves. Attention and learning are highly interdependent. Typically, students with attentional difficulties also have learning difficulties. According to several instructors and professional experts, students' attention spans are decreasing \cite{balan2020influence}. To address this issue, Mohammed Serrhini et al. evaluated students' attention in online education. During the learning process, students' attention is maintained by an EEG-based attention evaluation system. Attention data is maintained in a database and used by signal-processing algorithms to comprehend student knowledge growth. They concluded that BCI can be used to enhance learning levels in computer-based education\cite{serrhini2017toward}.

The use of manual sign systems is a means of communication between students with learning difficulties and their teachers. Due to a lack of learning support resources, instructors experience several practical obstacles while instructing children in manual sign language. To address these concerns, Youjin Choi et al. teamed up with instructors to design the Sondam Rhythm Game, a gesture-based rhythm game that aids the instruction of manual sign language. They conducted a four-week study with five teachers and eight students with learning difficulties. Based on video annotation and interviews, their game-based method to teach manual sign language has shown significant results. Their method increased children's attention span and motivation, as well as the number of spontaneous motions performed without prompting. In order to enhance teaching paradigms for eight students with learning difficulties, additional practical concerns and learning obstacles were identified. Based on their outcomes, they concluded that their proposed model method could be used to help learners improve their sign language abilities \cite{choi2022we}.

In summary, by enabling students to engage with technology using gestures, touch, and even brain signals, HCI can assist students in overcoming their learning challenges \cite{haraty2019associating}. Personalised user interfaces and design issues continue to be challenging issues for HCI domain.

\subsection{Internet of Things}
A system for connecting computers, mechanical and digital devices, things, or people with distinctive identities and the ability to transfer data via a human-to-human or computer-to-human interface is known as the Internet of Things (IoT). Both computers and humans can use IoT devices. IoT devices can transfer data over a network \cite{laghari2021review}. The IoT is considered by some to be a transformative force in education. The application of the digital technologies is not just making education omnipresent but also it is about making conventional systems of education more efficient and inclusive. IoT plays an important role in making education more interactive, collaborative, and accessible to all. IoT devices affords students reliable access to all learning resources, communication channels, and helps teachers keep track of student learning and progression in real-time \cite{wang2022smart}. Smart boards may be used in a similar manner to a blackboard for writing with a marker and can also show topic-related visuals and images to students. Global Positioning System (GPS) tracker-equipped school buses, smart security cameras, cellphones, and tablets with instructional applications will alter how schools and educational institutions have traditionally functioned. Teachers are concerned about student attendance, which is a daily requirement in schools. IoT can provide a solution to tracking and analysing student attendance for several purposes. Students are more engaging in virtual classrooms facilitated by smartphone applications. When they are able to comprehend more and more clearly, as described above, they are also able to think outside the confines of the classroom and to communicate and express their learning and questions \cite{verdes2021mobile}. The integration of IoT tools and smart devices will enable the educational curriculum is being changed and classroom surroundings are being made sound and light-sensitive to accommodate children with sensory disorders. In academia, IoT sensors collect data and automatically propose academic topics of interest to students for future learning processes \cite{mrabet2017smart}. These qualities make it more practical, and accessible for the students, instructors, and parents. It is a well-known reality that a quick transition of teaching methods and techniques cannot be done, but these gadgets are being customised and updated with the necessary software gradually and over time.

According to statistics, ASD has resulted in serious learning disorders. To address this issue, Raja et al. used Raspberry Pi to construct a system for evaluating the effectiveness of a smart monitor in assisting students with autism to learn and enhance their quality of life. They suggested a framework to support students with autism by assisting them in making choices, responding by telling parents what they are interested in and identifying their needs. They implement and evaluate a novel IoT-based system for supporting learning and enhancing the quality of life for students with autism. They claimed that their system could assist students with autism in understand any subject \cite{raman2021iot}.

Anna Lekova et al. designed, produced, and experimentally validated a Speech and Language Therapy (SLT) system for students with communication impairments. Their approach is able to interact with the IoT in order to assist SLT services in many educational and social contexts. It can link various assistive devices, APIs, online services, and agents to meet the particular requirements of each student using the intervention. Node-RED is used to link a humanoid NAO-type robot, an Emotiv EPOC+ brain headset, an emotionally expressive EmoSan robot, and a Kinect depth sensor. It is a flow-based tool for visual programming without the need to write code, and it can operate locally or on the IoT. The proposed system is sufficiently broad to be adapted to various kinds of therapy and to enable additional assistive devices and cloud services \cite{lekova2022system}.

Permanent or temporary Vision impairments present a number of obstacles in the daily living of a student with a learning difficulties. A student with vision impairment may be unable to distinguish between colours, which is a crucial aspect of work in various sectors. Humayun Rashid et al.  developed a colour-detecting system for the visually impaired. They addressed two extremely crucial difficulties for visually impaired individuals overcoming obstacles and falling. The proposed system combines the most recent hardware components, including an improved central processing unit, sensors for IoT and cloud-based architecture that effectively detects colour and obstacles. Moreover, it also alerts visually impaired individuals about colour and obstacles in multiple languages. In the event of fall detection, this proposed system also transmits a fall notice to the caretaker of the visually impaired individual, which is a major component of this work.

In summary, IoT can transform conventional learning methods and help students overcome their learning difficulties \cite{saade2018voice}. Security, standards, and dependence on AI judgements are challenges associated with the use of IoT in providing support for students with specific learning difficulties.

\subsection{Digital Twins}

The idea of using "digital twins" comes from NASA's Apollo program, in which at least two identical space vehicles were built. This allowed engineers to replicate the conditions of the spaceship during the trip, and the vehicle that stayed on Earth was called the twin \cite{boyes2022digital}. In 2002, Michael Grieves was the first person to talk about a Digital Twin (DT) \cite{newrzella2022three}. Previous research on DT definitions has shown that each system is made up of two parts: the physical system and a virtual system that includes all of the physical system's knowledge. Siemens describes it "A digital twin is a digital copy of a real product or process that is used to study and predict how it will work in the real world." Throughout the life-cycle of a product, digital twins are used to predict, simulate, and improve the product and manufacturing system before investing in real prototypes and assets \cite{newrzella2022methodology}.

A DT can help teachers to understand their students better and reduce the integration problems of a students with learning difficulties into inclusive higher education. Students with learning difficulties may find it difficult if they are repeatedly challenged to modify their classroom behaviour. The teachers may not have a good understanding of these students' behaviour. If a DT is built for a student, it enables teachers to carry out analysis of behaviour on the DT to give student related insights and help teachers to best support the academic and behavioural outcomes for their students. DT's also enable students with learning difficulties to work on digitally depicted scenarios and prepare them for the challenges of the real world. The combination of DT and VR will assist students in overcoming the challenges they face that are a result of their learning difficulties.

In conclusion, DT offers significant potential to address educational and behavioural issues relevant to students with learning difficulties. The application of DTs as an assistive technology remains largely unexplored territory. A major challenge for the widespread application of DTs is the inaccurate representation of the twin, since the construction of the twin relies on many different technologies, including AI, IoT, among others, and any errors in those interconnected technologies will result in the definition of an incorrect twin.

\subsection{The Metaverse} 

The word "meta" is a Greek word that signifies "more complete" or "transcending" and "Verse" is an abbreviated form of "universe." In his famous science fiction novel 'Snow Crash' published in 1992, the idea of the metaverse was first introduced by Neal Stephenson, in which people use digital avatars to control and compete with one another in order to advance their position. The use of VR and AR equipment helps the metaverse become more widespread \cite{huynh2023blockchain}. The metaverse is often described as a collection of socially conscious 3D virtual worlds \cite{peukert2022metaverse}.

The metaverse is a near-term technology offering significant opportunities for affording inclusive higher education. Regardless of their actual location and learning difficulties in the metaverse, students and teachers can meet in the digital world using their virtual reality headsets, promoting inclusive higher education. This capability can improve the teaching of individuals with learning difficulties. Inclusive design of the metaverse is however crucial to long term adoption in a number of domains including education \cite{IM}. The metaverse with XR has endless possibilities, with a potential influence on higher education that is especially significant for students with learning difficulties \cite{chengoden2022metaverse}. An inclusive metaverse-based school would  allow teachers to not only speak about their discoveries but also demonstrate them in a 3D environment. Together with their classmates, students with learning difficulties can engage in serious questioning and use firsthand experiences to assist in their academic development. No longer will students with learning disabilities be forced to sit in a traditional classroom with nothing to do. They can instead learn alongside their peers due to the educational and social opportunities afforded by the metaverse \cite{brown2001advanced}.

The use of the metaverse in inclusive education offers several potential advantages over traditional models, allowing students with learning difficulties to experience historical sites or conduct dangerous experiments in a secure, virtual setting alongside their classmates \cite{harris2022methodology}. Moreover, the metaverse learning environments can encourage safety in a manner that traditional classrooms cannot. Educators will have total control over student interactions in the metaverse and will be able to prohibit bullying. Thus, students with learning difficulties may concentrate on their education without worrying about bullies or other disruptions \cite{chengoden2023metaverse}. Roblox, Minecraft, Decentraland, Sandbox, Axie Infinity and Fortnite are some of available the metaverse projects \cite{gadekallu2022blockchain}. 

In conclusion, the metaverse offers significant potential for supporting students with learning challenges since it is a hybrid of cutting-edge and established technologies. The standardisation and interoperability of multiple technologies enable the metaverse to present a significant opportunity as an assistive technology for students with learning difficulties.

\section{ Tools }
In this section, we discuss a variety of tools for assisting students with various learning difficulties. An overview of these tools are shown in Table \ref{tab:Tools}.

\label{Sec:Tools}
\begin{table*}[h!]
\tiny
\centering
\caption{An Overview of Tools for Assisting Students with Learning Difficulties}
\label{tab:Tools}
\begin{tabular}{|p{2cm}|p{3.25cm}|p{1.25cm}|p{5.5cm}|}
\hline

\textbf{Tool Name}& \textbf{Disability Assisted} & \textbf{AT}& \textbf{Functionality} \\\hline
KURZWEIL 3000&Dyslexia& AI&It enables Text-to-Speech voices are accessible in 18 dialects and languages.\\\hline
QTrobot&Autism&AI& It can teach communication, emotions, and social skills through the use of facial expressions, gestures, and games.\\\hline
ActiveMath&Dyscalculia&AI&It can assist students develop interactive courses depending on the student's goals, preferences, skills, and prior knowledge.\\\hline
Widex's Evoke &CAPD&AI&It learns from the listening preferences of users throughout the world, bringing the development toward improved hearing out of the laboratory and into the real world.\\\hline
Microsoft Immersive Reader&Dyslexia, Dysgraphia, CAPD, VPD, and NVLD& AI&It Enhance reading and writing for people of all ages and abilities.\\\hline
Grammarly&Dysgraphia, CAPD, VPD, and NVLD&AI&It recognise faults in writing and look for a suitable replacement.\\\hline
Google Glass&CAPD, VPD, and NVLD&IoT, XR and AI&It provides basic voice or vision commands for online and Internet engagement.\\\hline
Google Expeditions&ALL&XR& Itis essentially a collection of XR experiences and "field excursions" offered by Google.\\\hline
Merge Cube & ALL&XR& It makes possible to study and interact with manipulate 3D digital things.\\\hline
CoSpaces Edu& ALL&XR& It helps students to create and engage with interactive media content.\\\hline
Assemblr&ALL&XR&Assemblr will enable educators to construct 3D objects and scenarios for classroom.\\\hline
Narrator AR&Dysgraphia, CAPD, VPD, and NVLD&XR&Narrator AR helps students improve their handwriting.\\\hline
Augmented Classroom&ALL&XR& It lets teachers provide their students access to 3D augmented environment.\\\hline
Roblox&ALL&Metaverse&it is a resource for teaching students about computer programming,
animation, 3D design, and application development.\\\hline
Minecraft&ALL&Metaverse&It is a digital world designed to foster innovation, teamwork, and
problem-solving though gaming.\\\hline
VoiceOver&VPD and NVLD&AI&It can describe what's on the screen in synthetic voice or braille through a linked braille display.\\\hline
Voice Dream Reader&dyslexia&AI&It is a fully functional document manager that enhances the built-in text to speech functionality with additional personalization options.\\\hline
TouchCast Studio&ALL&AI&It is a application that provides students with all the tools they need to create interactive films with hyperlinked hotspots.\\\hline
Book Creator&ALL&AI&It isa blank canvas on which students may generate an ebook.\\\hline

\end{tabular}
\end{table*}
\subsection{KURZWEIL 3000}
It is educational software designed to assist children with reading difficulties at home, at school, or in the workplace. This freeware includes OpenDyslexic typeface and text magnification to improve the readability of text for dyslexic students. Its 31 Natural Text-to-Speech voices are accessible in 18 dialects and languages, allowing students to access the same materials as their classmates. For a more meaningful learning experience, a test-preparation toolbar is accessible to students who wish to build their own evaluation \cite{kurzweil3000}.

\subsection{QTrobot}
QTrobot is a small, expressive AI-enabled humanoid developed for use by therapists and teachers. Children with autism spectrum disorder are taught communication, emotions, and social skills through the use of facial expressions, gestures, and games \cite{qtrobot}.

\subsection{ActiveMath}
ActiveMath is a web-based AI-learning system that develops interactive (mathematics) courses depending on the student's goals, preferences, skills, and prior knowledge. The material is delivered in an XML-based, semantic manner. The required information is acquired from a knowledge base for each student, and the course is constructed based on educational guidelines. The student is then provided with the course using a conventional web browser. ActiveMath is distinguished by its incorporation of independent mathematical service systems. This facilitates exploratory learning, realistically complicated activities, and the acquisition of proof methods \cite{activemath}.

\subsection{Widex's Evoke}
Widex's EVOKE is the first hearing aid in the world to incorporate machine learning. Every day, it improves the audio experience of every user. In addition, it learns from the listening preferences of users throughout the world, bringing the development toward improved hearing out of the laboratory and into the real world \cite{widexpro}.

\subsection{Microsoft Immersive Reader}
The free AI tool Microsoft's Immersive Reader enhance reading and writing for people of all ages and abilities. It is integrated into Word, OneNote, Outlook, Office Lens, Microsoft Teams, Reading Progress, Forms, Flipgrid, Minecraft Education Edition, and the Edge web browser \cite{ImmersiveReader}.

\subsection{Grammarly}
Grammarly is a typing assistant that uses AI and the cloud to check for problems in spelling, grammar, punctuation, clarity, engagement, and delivery. AI is used to recognise faults and look for a suitable replacement. Additionally, it gives users the option to customise their language, tone, and context\cite{Grammarly}.

\subsection{Google Glass}
Google Glass will provide basic voice or vision commands for online and Internet engagement. Google Glasses can easily power mobile devices like smartphones or tablets and may be thought of as wearable computers in this fundamental sense. The basic identification tool for blind and visually impaired people using the Google Glass camera. This viewpoint makes it possible to describe Google Glass as an assistive technology\cite{glass}.

\subsection{Google Expeditions}
Google Expeditions is essentially a collection of Augmented and Virtual Reality experiences and "field excursions" offered by Google. Some of the 'expeditions' are more supported than others with lesson plans, supporting links, and background information \cite{artsandculture}.

\subsection{Merge Cube}
The Merge Cube makes it possible to study and interact with the digital world in a whole new manner by letting you manipulate 3D digital things. In addition to many other things, students may inspect a DNA molecule, research the Earth's core, dissect a virtual frog, hold and share their own 3D creations, and explore a galaxy in the palm of their hand \cite{mergeedu}.

\subsection{CoSpaces Edu}
Users of CoSpaces, a web-based XR tool, may create and engage with interactive media content. CoSpaces allows students to demonstrate their knowledge in fresh ways by building interactive virtual settings that might be simple or complex but are still user-friendly for beginners \cite{cospaces}.

\subsection{Assemblr}
Assemblr will enable educators to construct 3D objects and scenarios for classroom usage. Students will have a better learning experience while utilising the Assemblr application to engage in AR, and VR \cite{assemblrworld}.

\subsection{Narrator AR}
An augmented reality software, Narrator AR helps students improve their handwriting. When a child's handwritten name is scanned, the application superimposes an animation showing the name blasting off the paper in the form of a rocket or a rainbow unicorn trail. Using the Narrator AR Mobile application, students can make a connection with their writing as virtual letters are lifted from the paper \cite{narratorar}.

\subsection{Augmented Classroom}
The purpose of the CleverBooks Augmented Classroom (ARC) is to let teachers provide their students access engaging, interactive courses by delivering content in fully immersive, 3D augmented settings. When used in the classroom, ARC's 3D environment has been shown to boost student engagement and motivation, leading to better academic outcomes \cite{AC}.

\subsection{Roblox}
Roblox Studio is a free, family-friendly resource for teaching students about computer programming, animation, 3D design, and application development. Using Roblox Studio in the classroom boosts students' self-esteem by giving them a real-world platform on which to practice the scientific inquiry skills outlined in the Next Generation Science Standards (NGSS).
Students can travel across time and space in Roblox adventures that are set up for exploration, investigation, and experimentation. Students may experience and analyse scientific events, go back in time to ancient Rome, and even construct virtual robots to compete against one another in friendly or hostile team challenges \cite{roblox}.

\subsection{Minecraft}
The educational version of Minecraft is a digital world designed to foster innovation, teamwork, and problem-solving through gaming. Teachers worldwide can use Minecraft education edition to capture their students' interest in a wide range of disciplines and make abstract concepts more concrete \cite{minecraft}.

\subsection{VoiceOver}
The iOS operating system has a built-in screen reader called VoiceOver. It can describe what's on the screen in synthetic voice or braille through a linked braille display for the visually impaired \cite{apple}.

\subsection{Voice Dream Reader}
Voice Dream Reader is a fully functional document manager that enhances the built-in text-to-speech functionality with additional personalisation options. These options include masking to display only a small portion of the text, support for dyslexia-friendly fonts, fully customisable colours for word and sentence highlighting, and more. Documents may be imported from a number of sources, such as Google Drive, Dropbox, and Bookshare, a programme that provides students with qualifying reading difficulties with free access to books in accessible formats \cite{VoiceDream}.

\subsection{TouchCast Studio}
TouchCast Studio is a free iPad application that provides students with all the tools they need to create interactive films with hyperlinked hotspots. Students can utilise a variety of video applications to link to supporting research on the Internet, ask questions through polls, link to an accessible script of their film, and more. The application has many complicated features, like a built-in teleprompter and green-screen features that let students work themselves from different places. It also works with multiple iPhone cameras \cite{TouchCast}.

\subsection{Book Creator}
The Book Creator provides a blank canvas on which students may generate an ebook to demonstrate their comprehension and incorporate all of their media. Each book may have text, photographs with descriptive captions, audio, and video. The free, fully-functional version of Book Creator for iPad may be used to produce one book. Upgrade to the premium version to have access to limitless publishing and comic book templates. There is also a Chromebook-compatible web-based version of Book Creator; with this edition, users may generate up to 40 free booklets \cite{bookcreator}.

\section{Projects}
In this section, we discuss some related projects that work toward students with learning difficulties and inclusive education. The overview of the related projects that work towards students with learning difficulties and inclusive education is shown in Table \ref{tab:Comparision} 

\begin{table*}[h!]
\tiny
\centering
\caption{Overview of related projects that work towards the student with learning difficulties and inclusive education}
\label{tab:Comparision}
\begin{tabular}{|p{2.5cm}|p{2.5cm}|p{4cm}|p{2cm}|}
\hline

\textbf{Project Name}& \textbf{Key Action} & \textbf{Action Type} & \textbf{Countries covered}\\\hline
IncED&Partnerships for cooperation and exchanges of practices&Cooperation partnerships in youth& Latvia, Portugal, Estonia, and Spain\\\hline
Fairness Inclusive&Learning Mobility of Individuals&Youth mobility& Russia, Romania, France, and Germany\\\hline
InClusion Team& Cooperation for innovation and the exchange of good practices& Strategic Partnerships for school education& Portugal, Greece, and Spain\\\hline
ELLeN&Cooperation for innovation and the exchange of good practices&Strategic Partnerships for higher education& Germany, Austria, and Belgium \\\hline
Accessible Peer Interaction with Disabled Youth& Accessible peer interaction with disabled youth& To impact young people, thereby investing in the future and providing the proper basis for a spillover effect into wider society and into the working environment. &Bulgaria, Belgium, Austria, and United Kingdom\\\hline

\end{tabular}
\end{table*}
\subsection{IncED}
This project is managed by the private organisation Vivere kool MTU and funded by Erasmus+, the European Union's education, training, youth, and sports programme. This project focuses on inclusion, fostering equality and non-discrimination, and using modern teaching-learning techniques. The project's objectives are to better understand concepts of mixed-ability and inclusive collaboration and to highlight it as a possibility rather than a challenge, to adapt current systems and build innovative educational games and methodologies for collaborating in mixed-ability groups that can be used in non-formal and formal learning contexts for students aged 13 to 18 years old, to promote inclusive education among various types of stakeholders, and to improve the youth workforce \cite{IncED}.
\subsection{Fairness Inclusive}
This project is handled by the Kreisau Initiative association and funded by Erasmus+, the European Union's programme for education, training, youth, and sport. This project focuses on disability and special needs, equality and access for the underprivileged. One of the primary objectives of the initiative is to engage disadvantaged youth in international educational activities. This engagement help the emancipation of young people so that they may lead more independent lives and treat others with tolerance, solidarity, and respect in the future. The underlying belief of this objective is that it increases their social inclusion and minimises their marginalisation \cite{Fairness}.
\subsection{InClusion Team}
 This project is managed by Fyllingsdale High School in Norway. The partners include universities, secondary schools, a public teacher training centre, and educationally focused non-governmental organisations. Erasmus+, the European Union's programme for education, training, youth, and sport, is funding this project. This project focuses on new technologies, digital skills, access for the disadvantaged, and particular requirements for those with impairments. The project aims to establish a learning community in which universities, schools, teacher training centres, and non-governmental organisations share best practices for teaching ICT to students and individuals with special needs. The collaborating parties intend to equip educators with training and expand access to high-quality learning tools, and materials \cite{InClusion}.

\subsection{ELLeN}
This project is managed by Goethe University Frankfurt and funded by Erasmus+, the European Union's education, training, youth, and sport programme. In this project, a teacher will teach students how to research the needs of a certain group of learners by interviewing those learners. In other words, neurodiverse learners will be treated as experts in their own learning process. By helping pre-service teachers build up their Inquiry-based learning skills and competencies, this initiative also shows how to partnership with neurodiverse learners, the strengths and needs of the target population's learners can be identified, evaluated, and met in learning environments.\cite{ELLeN}.

\subsection{Accessible Peer Interaction with Disabled Youth}
The National Association of Professionals Working with People with Disabilities manages this project (NARHU, Bulgaria). This initiative is supported by Erasmus+, the education, training, youth, and sports programme of the European Union. This project targets youth workers and leaders, student leaders, student bodies, youth organisation leaders, and representatives of disabled youth organisations. The work aims to assist youth workers in developing and disseminating effective strategies for reaching out to marginalised youth, refugees, asylum seekers, and migrants, as well as combating racism and intolerance among young people \cite{peer}.

\section{ Challenges and Future Directions}
\label{Sec:Challenges}
\begin{figure*}[ht!]
	\centering
	\includegraphics[width=\linewidth]{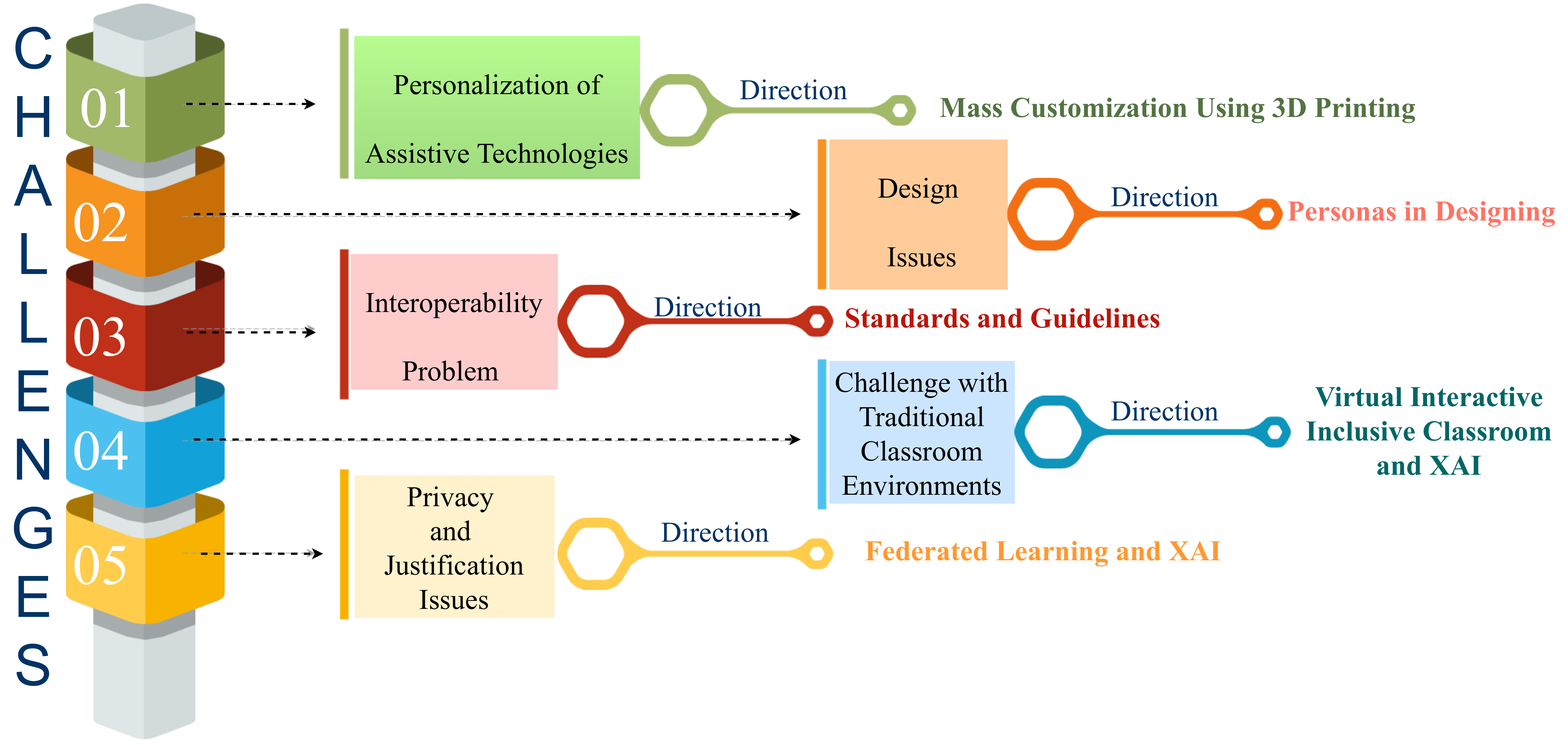}
	\caption{Challenges and Future Directions of Assistive Technologies for Students with Learning Difficulties in Higher Education}
	\label{CF}
\end{figure*}

This section discusses challenges and future directions of assistive technologies for students with learning difficulties in higher education. An overview of these challenges and future directions is depicted in Fig \ref{CF}.

\subsection{Personalization of Assistive Technologies} 
\textbf{Challenge:} It is not possible for a single assistive technology to represent a universal solution to the learning challenges faced by all learning difficulties. The need for these assistive technologies varies with the needs and preferences of each student with learning difficulties. Some students may or may not need the full functionality of a particular assistive technology, while others may require a combination of the functionality of many assistive technologies. Personalisation of these assistive technologies is very important for inclusive education, but which still remains a challenge that must be addressed.

\textbf{Potential Solution:} Students with learning difficulties may customise their assistive technology to meet their needs in a virtual world by using automated 3D modelling and printing. The organisations must also permit mass customisation of educational aids that help students overcome their learning difficulties \cite{stramandinoli2022maxillofacial}.

\subsection{Design Issues}

\textbf{Challenge:} A variety of computer interfaces and technologies will be used for academic purposes by students without learning difficulties. While individuals with learning difficulties need a personalised environment, this may or may not be available in their classroom environments. If the instructor does not provide such assistive technology to the students, the inclusive education concept would be compromised. The challenge with HCI still exists, and it not feasible for students with learning difficulties to use the same computer interfaces and technology as effectively as their classmates.

\textbf{Potential Solution:}
The usage of personas in designing an HCI takes into account a variety of characters with and without learning difficulties, which aids in the better design of services and products that can be utilised by everyone regardless of any  difficulties \cite{salminen2022persona}.

\subsection{Interoperability Problem among Assistive Technologies}
\textbf{Challenge:} The devices and technologies used in the creation of various assistive technologies may not be developed by a single organisation. There are currently no interoperability standards or regulations in place for assistive technology like the metaverse. As a result, these assistive technologies will raise integration and interoperability challenges. This could result in the student losing focus in an inclusive educational environment.

\textbf{Potential Solution:} Instead of only providing best practices, there is a need for standards that address these issues relating to the creation of assistive technologies. These guidelines need to specify the industry standards for developing assistive technologies. Furthermore, verification and certifications from the scientific community are required before an assistive technology is delivered to the market. There should be specific guidelines for utilising them while adapting to the real world. This will allow enterprises to build assistive devices more quickly and at a lower cost. Additionally, this will also solve the interoperability problems with different assistive technologies \cite{elsaesser2022standard}.

\subsection{Challenge with Traditional Classroom Environments}
\textbf{Challenge:} Traditional classroom environments may have a negative impact on all students to some extent, at some time, and in some way, but students with learning difficulties are more vulnerable. This difficulty stems from their difficulties with speaking, writing, or thinking. Incorporating assistive technology into a conventional classroom environment exacerbates the challenge by requiring students to bring gadgets to class and use them to overcome their difficulties. These students' gadgets are unusual among their classmates, and if any technical difficulties arise, they further exacerbate their problems.

\textbf{Potential Solution:} A Virtual Interactive Inclusive Classroom (VIIC) can help students with learning difficulties overcome the challenges they face in traditional classroom settings. Students can interact with one another in this class while taking classes from a convenient location. The virtual interactive, inclusive classrooms will use cutting-edge technology such as Explainable Artificial Intelligence (XAI) to help students make decisions, sophisticated Mixed Reality (MR) for in-person interactions, and advanced robotic assistance that can be linked with the brain of a student with a learning difficulties to help them do academic tasks. Thus, VIIC can help students overcome their disabilities and compete with their peers \cite{maddikunta2022industry}.

\subsection{Privacy and Justification Issues}
\textbf{Challenge:} AI can be combined with other technologies, such as digital twins, IoT, HCI, and the metaverse, to produce better outcomes. Students with learning difficulties must share personal data to receive decision-making assistance from AI models that support other technologies. These AI models cannot provide privacy for data provided by students with learning difficulties, which may create distrust of assistive technologies. The recommendation from AI-based assistive technologies cannot be completely depended on as they are black box in nature \cite{webb2021machine}.

\textbf{Potential Solution:} Federated learning can help with the problem of data privacy in assistive technologies. Federated learning addresses data ownership and privacy by guaranteeing that data never leaves dispersed node devices. Simultaneously, the global model is updated and distributed to all network nodes. This ensures the privacy of the students' data \cite{banabilah2022federated}. XAI can help AI-based decisions more justifiable and accountable. XAI is a set of procedures and techniques that allow users to understand and rely on the outcomes. XAI will increase the accountability of recommendations. XAI can explain the anticipated outcomes and any potential biases. It aids in describing a suggestion's accuracy, fairness, and transparency and improves decision-making. XAI increases trust and confidence in assistive technology suggestions \cite{dwivedi2022explainable}. 

\section{Roadmap}
Based on the findings of the systematic literature review that we carried out, AI, XR, IoT, HCI, digital twins, and the metaverse are the potential technologies that can assist students with learning difficulties. In the past few decades, AI and HCI have played a significant role in supporting students who struggle with a variety of learning difficulties. As research progresses, AI and HCI as forms of assistive technology will continue to develop. IoT as an assistive technology enhances the voice and vision and also provides real-time data on various challenges faced by students with learning difficulties using sensors. AR and VR, as assistive technologies, will effectively assist students with learning difficulties in engaging in activities and social integration. Even though the IoT, AR, and VR are already being used as assistive technologies in developed countries, it will take a significant amount of time for these technologies to catch on in developing or underdeveloped countries. Assistive technologies like digital twins require a huge number of IoT sensors to work effectively, and these digital twins, which are replicas of real-world objects, can help experts mitigate the challenges of learning difficulties virtually before being applied to the real world. The metaverse as assistive technology, which is enabled by various technologies like blockchain, edge computing, quantum computing, 3D modelling, XR, IoT, 6G, AI, and others, will make students with learning difficulties equivalent to their peers. The digital twins and the metaverse will be the future of assistive technologies that will help students overcome the challenges raised by their learning difficulties.


\section{Conclusion }
\label{Sec:Conclusion}
This survey aims to understand the potential role of a range of recent developments in technology in providing inclusive education for students with learning disorders in higher education. Throughout this process, we have analysed the learning and support needs that arise as a result of a student's learning difficulties and how these recently technologies developed can support inclusive education for students with learning difficulties. We have searched a range of online digital libraries to locate journal articles and conference papers relevant to our study. According to our examination of the relevant literature, there has been little research on assistive technologies related to XR, IoT, digital twins, and the metaverse. The use of AI and HCI as assistive technologies is more prevalent than other technologies. It is also understood that research on aiding students with learning difficulties in primary education is more widespread than in higher education. The selection of these technologies within this review is also supported by recent reports identifying them as important developments in educational technology for higher education in the near, mid and longer term future. Our review has also highlighted projects related to assistive technology for the inclusive education students with learning difficulties in higher education. We also proposed strategies that may aid individuals with learning difficulties in higher education. Moreover, we highlighted the challenges of using assistive technology in providing inclusive education for students with learning difficulties in higher education and providing potential solutions. We aim to provide a road map to describe an accessible and inclusive higher education system using important and highly promising technologies to support students with learning difficulties.


\bibliography{Ref}
\bibliographystyle{IEEEtran}

\begin{IEEEbiography}[{\includegraphics[width=1in,height=1.25in,clip,keepaspectratio]{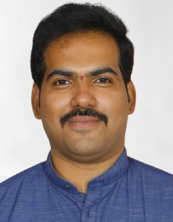}}]{Gokul Yenduri}received his Master’s degree from Vellore Institute of Technology in the year 2013. Currently, he is a senior research fellow at the DIVERSASIA project, co-funded by the Erasmus+ programme of the European Union. His areas of interest are Metaverse, Inclusive Education, Federated Learning, Blockchain, and Explainable AI. He has worked as an assistant professor in the past. He attended several national and international conferences, workshops, and guest lectures and published papers in peer-reviewed international journals. He is also acting as a reviewer for many prestigious peer-reviewed international journals.
\end{IEEEbiography}
\begin{IEEEbiography}[{\includegraphics[width=1in,height=1.25in,clip,keepaspectratio]{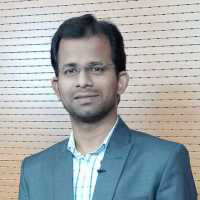}}]{Dr. Rajesh Kaluri} is currently working as an Associate Professor in the School of Information Technology and Engineering, Vellore Institute of Technology, Vellore, Tamil Nadu, India. He has more than ten years of teaching experience. He was a visiting Professor in Guangdong University of Technology, China in 2015 and 2016. He has completed a Ph.D. in Computer Vision at VIT University, India. He obtained M. Tech in CSE from ANU, Guntur, India, and B. Tech in CSE from JNTU, Hyderabad. His current research is in the areas of Human Computer Interaction, Computer Vision, Big data Analytics, IoT. He is partnering with the Nottingham Trent University in the United Kingdom on a funded project worth 80 lakhs which is financed by the Erasmus + Programme of the European Union. He is associated with several reputed journals as a guest editor. He has published 40+ research papers in various reputed international journals. 
\end{IEEEbiography}
\begin{IEEEbiography}[{\includegraphics[width=1in,height=1.25in,clip,keepaspectratio]{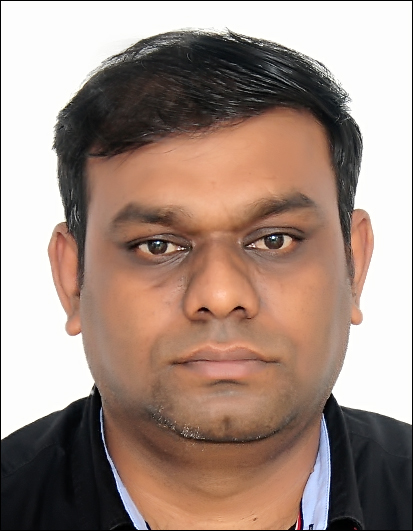}}]
{Dr. Dharmendra Singh Rajput} working as an Associate Professor in the Department of Software and Systems Engineering, SITE, VIT, Vellore since June 2014. He has completed a Ph.D. (2014) from NIT, Bhopal, India. His research areas are data mining, machine learning, and big data. He has published 35+ reputed Journal Papers, 5 edited books published under reputed publishers, and 17 papers presented in the reputed international conference. He is also a guest editor of various reputed journals. He has received various awards from the Indian Government like DST-SERB, CSIR Travel Grant, and MPCST Young Scientist Fellowship. He is doing the funded project of 80 lakhs which is received from Erasmus + Programme of the European Union with the partner the University of Nottingham UK. He has visited various countries UK, France, Singapore, UAE, China, and Malaysia for academic purposes.
\end{IEEEbiography}
\begin{IEEEbiography}[{\includegraphics[width=1in,height=1.25in,clip,keepaspectratio]{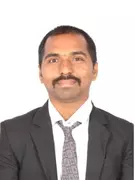}}]
{Dr. Kuruva Lakshmanna} has received his B-Tech in Computer Science and Engineering from Sri Venkateswara University College of Engineering -Tirupathi, India in the year 2006, M-Tech in Computer Science and Engineering (Information Security) from National Institute of Technology Calicut, Kerala, India in the Year 2009, and Ph.D from VIT, India in the year of 2017. I am working as an Associate professor in VIT, India. I was a visiting professor in Guangdong University of Technology, China in 2018. My research interests are Machine learning, Data Mining in DNA sequences, IoT, Block chain etc. I have more than 12 years of experience in teaching. I have published around 25 papers in various reputed interantional journals. I am an editor in Hindawi journals. He is doing the funded project of 80 lakhs which is received from Erasmus + Programme of the European Union with the partner the University of Nottingham UK.
\end{IEEEbiography}

\begin{IEEEbiography}[{\includegraphics[width=1in,height=1.25in,clip,keepaspectratio]{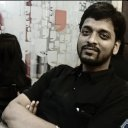}}]{Dr. Thippa Reddy Gadekallu} is currently working as an Associate Professor in the School of Information Technology and Engineering, Vellore Institute of Technology, Vellore, Tamil Nadu, India. He obtained his Bachelors in Computer Science and Engineering from Nagarjuna University, India, in the year 2003, Masters in Computer Science and Engineering from Anna University, Chennai, Tamil Nadu, India in the year 2011 and his Ph.D. in Vellore Institute of Technology, Vellore, Tamil Nadu, India in the year 2017. He has more than 14 years of experience in teaching. He has more than 100 international/national publications in reputed journals and conferences. Currently, his areas of research include Machine Learning, Internet of Things, Deep Neural Networks, Blockchain, Computer Vision. He is an editor in several publishers like Springer, Hindawi, Plosone, Scientific Reports (Nature), Wiley. He also acted as a guest editor in several reputed publishers like IEEE, Springer, Hindawi, MDPI. He is recently recognized as one among the top 2\% scientists in the world as per the survey conducted by Elsevier in the year 2021.
\end{IEEEbiography}

\begin{IEEEbiography}[{\includegraphics[width=1in,height=1.25in,clip,keepaspectratio]{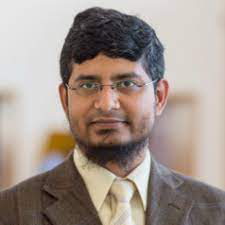}}]
{Dr. Mufti Mahmud} is an Associate Professor of Cognitive Computing at the Computer Science Department of Nottingham Trent University (NTU), UK. He has been listed among the Top 2\% cited scientists in the world for 2020. He was also the recipient of the NTU VC outstanding research award 2021 and the Marie-Curie postdoctoral fellowship. Dr Mahmud is the coordinator of the Computer Science and Informatics research excellence framework unit of assessment at NTU and the deputy group leader of the Cognitive Computing \& Brain Informatics and the Interactive Systems research groups. His research portfolio consists of GBP3.3 million grant capture with expertise that includes brain informatics, computational intelligence, applied data analysis, and big data technologies focusing on healthcare applications. He has over 15 years of academic experience and over 200 peer-reviewed publications. Dr Mahmud is the General Chair of the Brain Informatics conference 2020, 2021, and 2022; Applied Intelligence and Informatics conference 2021 and 2022; chair of the IEEE CICARE symposium since 2017 and was the local organising chair of the IEEE WCCI 2020. He is the Section Editor of the Cognitive Computation, the Regional Editor (Europe) of the Brain informatics journal and an Associate Editor of the Frontiers in Neuroscience. During the year 2021-2022, Dr Mahmud has been serving as the Vice-Chair of the Intelligent System Application and Brain Informatics Technical Committees of the IEEE Computational Intelligence Society (CIS), a member of the IEEE CIS Task Force on Intelligence Systems for Health, an advisor of the IEEE R8 Humanitarian Activities Subcommittee, the Publications Chair of the IEEE UK and Ireland Industry Applications Chapter, and the Project Liaison Officer of the IEEE UK and Ireland SIGHT Committee, the Secretary of the IEEE UK and Ireland CIS Chapter, and the Social Media and Communication Officer of the British Computer Society's Nottingham and Derby Chapter.
\end{IEEEbiography}

\begin{IEEEbiography}[{\includegraphics[width=1in,height=1.25in,clip,keepaspectratio]{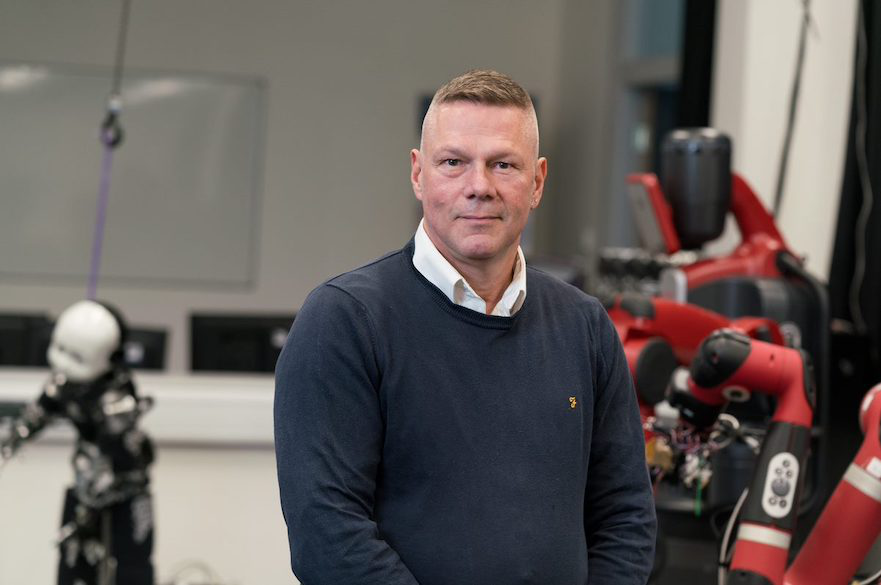}}]
{Dr. David J Brown} is currently a professor of interactive systems for social inclusion with Nottingham Trent University. He is the principal investigator for two EU H2020 Grants (MaTHiSiS and No One Left Behind) to investigate the use of sensor data to understand the emotional state of learners to provide personalised learning experiences, and how game making can enhance the engagement of students with learning disabilities and autism. He is also co-investigator to EPSRC Internet of Soft Things, to investigate the impact of networked smart textile objects on young people's wellbeing. 
\end{IEEEbiography}

\EOD

\end{document}